\documentclass[aps,showpacs,amsmath,amssymb]{revtex4}

\usepackage{graphicx}
\usepackage{dcolumn}
\usepackage{bm}
\usepackage{hyperref}

\newcommand{\beq}{\begin{equation}}
\newcommand{\eeq}{\end{equation}}
\newcommand{\beqa}{\begin{eqnarray}}
\newcommand{\eeqa}{\end{eqnarray}}
\newcommand{\beqar}{\begin{eqnarray*}}
\newcommand{\eeqar}{\end{eqnarray*}}
\newcommand{\bra}[1]{\mbox{$\left\langle{#1}\right|$}}
\newcommand{\ket}[1]{\mbox{$\left|{#1}\right\rangle$}}

\newcounter{saveeqn}

\begin{document}

\title{A case concerning the improved transition probability}
\author{Jian Tang}
\affiliation{Quantum Theory Group, Department of Modern Physics,
University of Science and Technology of China, Hefei, 230026,
P.R.China}
\author{An Min Wang}\email{anmwang@ustc.edu.cn}
\affiliation{Quantum Theory Group, Department of Modern Physics,
University of Science and Technology of China, Hefei, 230026,
P.R.China}

\begin{abstract}

As is well known, the existed perturbation theory can be applied to
calculations of energy, state and transition probability in many
quantum systems. However, there are different paths and methods to
improve its calculation precision and efficiency in our view.
According to an improved scheme of perturbation theory proposed by
[An Min Wang, quant-ph/0602055 v7], we reconsider the transition
probability and perturbed energy for a Hydrogen atom in a constant
magnetic field. We find the results obtained by using Wang's scheme
are indeed more satisfying in the calculation precision and
efficiency. Therefore, Wang's scheme can be thought of as a powerful
tool in the perturbation calculation of quantum systems.

\end{abstract}


\maketitle

\section{Introduction}\label{sec1}

The traditional method of perturbation theory tells us to calculate
the energy and expansion coefficient step by step. That is, we
should first calculate the zeroth order energy and wave function,
and then the first,the second, and so on. In fact, it is easy to see
that such a way introduces the approximation too early, and each
order calculation of energy and expansion coefficient of wave
function are based on the result of the former orders. After careful
examination, we find that the traditional way does not consider the
astringency of the expansion of the wave function. There is every
possibility that if we reconsider the astringency, the result might
be different. In fact, this is effectively embodied in Wang's scheme
of perturbation theory \cite{wang1}. In this reference, the author
did not introduce approximation until very late, and consider subtly
and systemically the affection of high-order approximation to the
low-order one. This finally results in a different formalism of
expression of perturbed solution of dynamics, and its expansion
coefficients contain reasonably the high-order energy amendment.

It can been seen that some physics expressions were modified
according to Ref. \cite{wang1} and further applications can be
expected. This leads us to think that it is important and
interesting to consider their influences on the physical problems
via first reexamining those familiar and standard examples, and then
studying more practical systems, because better precision, higher
efficiency, as well as correct physical features are always the aims
that physics pursues continuously. Here, we focus our attention on a
typical example, which shows satisfying results in the calculation
precision, efficient as well as the corresponding physical features.
This implies, in our view, that Wang's scheme of perturbation theory
is a powerful tool in the perturbative calculation of quantum
system.

In this paper, we intend to study the revisions on transition
probability in Ref. \cite{wang1}, which is different from
traditional one. We find that Wang's revision to the existed
expression of transition probability is never trivial, and we
illustrate this conclusion via calculating this revision for a
Hydrogen atom in a constant magnetic field. This example concerns
the ground state hyperfine structure of Hydrogen atom whose
correction of electron self-energy has been studied (see
\cite{self}), so does similar problem about muonium (see \cite{nio}
and \cite{liu}). By comparing our result of transition probability
with traditional one, our above view is verified. After referring to
the exact solution of this problem, we will see that the improved
transition probability given by \cite{wang1} does show some of its
advantages. From this case it is necessary to realize that perhaps
there are also some other problems that need similar revisions in
transition probability. We are sure that applications of Wang's
scheme of perturbation theory to other problems should not be
unimportant and short of practical significance, although his scheme
just contains the higher order revisions.

To effectively organize this article, we divide it into the
following main parts. Besides Sec. \ref{sec1} which is an
introduction, in Sec. \ref{sec2} we first introduce the amendment to
the transition probability based on the results of \cite{wang1}.
Next, in Sec. \ref{sec3} we provide the calculation of energy of our
example according to the Wang's scheme, then compare it with the
exact result, and these results are helpful for us later to
calculate the transition probability. Then, in Sec. \ref{sec4} we
will take use of Sec.\ref{sec2} and Sec.\ref{sec3} to calculate the
referred case. Finally, in Sec. \ref{sec5} we summarize our
conclusions and make some discussions.

\section{Improved TRANSITION PROBABILITY}\label{sec2}

Let us start with Wang's scheme of perturbation theory \cite{wang1},
denoting the state vector by \beq \label{inis1}
\ket{\Psi(t)}=\sum_{l=0}^{\infty}\sum_{\gamma}c_{\gamma,I}^{(l)}(t)\ket{\phi^{\gamma}},
\eeq where $\ket{\Psi(t)}$ is the eigenvector of ${H}$, and
$\ket{\phi^{\gamma}}$ is the eigenvector of ${H_0}$. According to
the improved form of the perturbed solution of dynamics in
\cite{wang1}, the first order amendment of coefficient of state
vector is deduced as \beq
c_{\gamma,I}^{(1)}(t)=\frac{g_{1}^{\gamma\beta}}
{E_{\gamma}-E_{\beta}}\left(1-e^{i\widetilde{\omega}_{\gamma\beta}t}
\right), \eeq where $E_{\gamma}$ are eigenvalues of $H_0$ with the
eigenvector $\ket{\phi^\gamma}$, $g_{1}^{\gamma\beta}$
($\gamma\neq\beta$) are off-diagonal elements of $H_1$ in the
representation of $H_0$, that is
$g_1^{\gamma\beta}=\bra{\phi_\gamma}H_1\ket{\phi_\beta}$, and \beq
\widetilde{\omega}_{\gamma\beta}=\widetilde{E_{\gamma}}-\widetilde{E_{\beta}},
 \eeq while $\widetilde{E_{\gamma}}$ is so-called improved form of
perturbed energy defined by\beq
\widetilde{E_{\gamma}}=E_{\gamma}+h_1^{\gamma}
+G_{\gamma}^{(2)}+G_{\gamma}^{(3)}+G_{\gamma}^{(4)}+\cdots. \eeq
Here, $h_1^\gamma$ are diagonal elements of $H_1$ in the
representation of $H_0$, and \beq
G_{\beta}^{(2)}=\sum_{{\beta}_{1}}\frac{1}{E_{\beta}-E_{{\beta}_{1}}}
g_{1}^{\beta{\beta}_{1}}g_{1}^{{\beta}_{1}\beta}, \eeq  \beq
 G_{\beta}^{(3)}=\sum_{{\beta}_{1},{\beta}_{2}}
\frac{1}{(E_{\beta}-E_{{\beta}_{1}})(E_{\beta}-E_{{\beta}_{2}})}
g_{1}^{\beta{\beta}_{1}}g_{1}^{{\beta}_{1}{\beta}_{2}}g_{1}^{{\beta}_{2}\beta},
\eeq \beq G_{\beta}^{(4)}=\sum_{{\beta}_{1},{\beta}_{2},{\beta}_{3}}
\frac{g_{1}^{\beta{\beta}_{1}}g_{1}^{{\beta}_{1}{\beta}_{2}}
g_{1}^{{\beta}_{2}{\beta}_{3}}g_{1}^{{\beta}_{3}\beta}
\eta_{\beta{\beta}_{2}}}{(E_{\beta}-E_{{\beta}_{1}})
(E_{\beta}-E_{{\beta}_{2}})(E_{\beta}-E_{{\beta}_{3}})}\\
-\sum_{{\beta}_{1},{\beta}_{2}}\frac{g_{1}^{\beta{\beta}_{1}}
g_{1}^{{\beta}_{1}\beta}g_{1}^{\beta{\beta}_{2}}
g_{1}^{{\beta}_{2}\beta}}{{(E_{\beta}-E_{{\beta}_{1}})}^{2}(E_{\beta}-E_{{\beta}_{2}})},
\eeq where $\eta_{\beta{\beta}_{2}}=1-\delta_{\beta{\beta}_{2}}$,
and $\eta_{\beta{\beta}_{2}}=0$, if $\beta={\beta}_{2}$;
$\eta_{\beta{\beta}_{2}} \neq 0$, if $\beta \neq {\beta}_{2}$.
Thus it leads to so-called improved transition probability that is
different from the traditional one, that is \beq
\label{newprb}P_{I}^{\gamma\beta}(t)=\left|g_{1}^{\gamma\beta}\right|^{2}
\frac{\sin^{2}(\widetilde{\omega}_{\gamma\beta}t/2)}{(\omega_{\gamma\beta}/2)^{2}}.
\eeq

\section{Revision of perturbed ENERGY}\label{sec3}

Now we begin to study such a system in which a Hydrogen atom is
placed in a uniform constant magnetic field, which is in the $+z$
direction in the rectangular coordinate system. Suppose that the
atom is at the ground state, thus its Hamiltonian (see Ref.
\cite{codata} Appendix(D1)) is \beq {H}= {H_{0}} + {H_{1}}
\\  = B(\mu_{e}\sigma_{ez} -\mu_{p}\sigma_{pz})+
W{{\bm{\sigma}}_{e}}\cdot {{\bm{\sigma}}_{p}}. \eeq In the equation
above, we change some symbols that are different from the equation
(D1) in Ref. \cite{codata}. $B$ is the magnitude of magnetic field.
$W$ is a coupling constant, and we will give its value later in Sec.
\ref{sec4}. $\mu_{e}$ is the electron magnetic moment. $\mu_{p}$ is
the magnetic moment of a proton, and ${\bm{\sigma}} =
(\sigma_{x},\sigma_{y},\sigma_{z} )$ is the Pauli operator. Since it
is well known that $\mu_{p} \ll \mu_{e}$, we can easily omit the
$B\mu_{p}\sigma_{pz}$ in the ${H}$. Therefore, the division of
Hamiltonian can be written as \beq \label{h}{H}= {H_{0}} + {H_{1}} =
B\mu_{e}\sigma_{ez} + W{{\bm{\sigma}}_{e}}\cdot{{\bm{\sigma}}_{p}}.
\eeq Here, ${H_{0}}=W{{\bm{\sigma}}_{e}}\cdot{{\bm{\sigma}}_{p}}$ is
the unperturbed part, and ${H_{1}}= B\mu_{e}\sigma_{ez}$ is taken as
the perturbed part.

Now we refer to the way proposed by Wang in Ref. \cite{wang1}.
First, we should calculate the eigenvalues and eigenvectors of
${H_{0}}$. It is easy since we know that ${H_{0}}$ is just the
coupling of spins of electron and proton. Here we just list the
results and omit the process of calculating them. \beq
\quad\mbox{Eigenvalue:} \quad W \Longrightarrow
\quad\mbox{Eigenvectors:}\quad \left\{
\begin{array} {r@{\quad = \quad}l} \phi_{1}& \alpha(e)\alpha(p)\\[8pt] \phi_{2}&
\frac{1}{\sqrt{2}}[\alpha(e)\beta(p)+ \beta(e)\alpha(p)]\\[8pt] \phi_{3}&
\beta(e)\beta(p) \end{array} \right. ,\eeq and \beq
\quad\mbox{Eigenvalue:} \quad -3W \Longrightarrow
\quad\mbox{Eigenstate:}\quad \phi_{4}
\,=\,\frac{1}{\sqrt{2}}[\alpha(e)\beta(p)-\beta(e)\alpha(p)] \eeq
Here, $\alpha$ and $\beta$ mean: \beq \label{alpha}
\alpha=\begin{pmatrix} 1\\
0 \end{pmatrix} \eeq   \beq \label{beta}
\beta=\begin{pmatrix} 0\\
1 \end{pmatrix} \eeq It is clear that there is a three-fold
degeneracy subspace.

Considering that \beq
\sigma_{ez}\ket{\phi_{1}}=\ket{\phi_{1}},\quad\sigma_{ez}\ket{\phi_{2}}
=\ket{\phi_{4}},\quad
\sigma_{ez}\ket{\phi_{3}}=-\ket{\phi_{3}},\quad\sigma_{ez}\ket{\phi_{4}}=\ket{\phi_{2}}
\eeq The ${H}$ matrix in the representation of $H_0$ can be written
as following : \beq {H}=\begin{pmatrix}
W+B\mu_{e}&0        &0         & 0 \\
0         &W        &0         & B\mu_{e}\\
0         &0        &W-B\mu_{e}& 0\\
0         &B\mu_{e} &0         &-3W \end{pmatrix} \eeq
\\
According to Ref. \cite{wang1} , we should redivide ${H}$ into two
parts: the diagonal part ${H_{0}^{\prime}}$ and the off-diagonal
part ${g_{1}}$, that is, \beq {H_{0}^{\prime}}=\begin{pmatrix}
W+B\mu_{e}&0        &0         & 0 \\
0         &W        &0         & \\
0         &0        &W-B\mu_{e}& 0\\
0         &0        &0         &-3W
\end{pmatrix} \eeq
\beq  {g_{1}}=\begin{pmatrix}
0         &0        &0         & 0 \\
0         &0        &0         & B\mu_{e}\\
0         &0        &0         & 0\\
0         &B\mu_{e} &0         & 0 \end{pmatrix} \eeq It is clear
that the degeneracy is completely removed by the redivision skill
\cite{wang1}.

Again from Ref. \cite{wang1}, we have \beq
\widetilde{E_{\beta}}=(H_0^\prime)_{\beta\beta}+G_{\beta}^{(2)}+G_{\beta}^{(3)}+G_{\beta}^{(4)}+\cdots
\eeq Note that $(H_0^\prime)_{\beta\beta}=E_\beta+h_1^{\beta}$.
Thus, using the given matrix above, it is easy to obtain that: \beq
G_{1}^{(2)}=G_{1}^{(3)}=G_{1}^{(4)}=0, \eeq \beq
G_{2}^{(2)}=\frac{{(B{\mu}_{e})}^{2}}{4W},\quad G_{2}^{(3)}=0, \quad
G_{2}^{(4)}=-\frac{{(B{\mu}_{e})}^{4}}{(4W)^3}\eeq \beq
G_{3}^{(2)}=G_{3}^{(3)}=G_{3}^{(4)}=0,\eeq \beq
G_{4}^{(2)}=-\frac{{(B{\mu}_{e})}^{2}}{4W},\quad G_{4}^{(3)}=0,
\quad G_{4}^{(4)}=\frac{{(B{\mu}_{e})}^{4}}{(4W)^3}\eeq Thus \beqa
\widetilde{E_{1}}&=& W+B{\mu}_{e},\\
\widetilde{E_{2}}&=&
W+\frac{{(B{\mu}_{e})}^{2}}{4W}-\frac{{(B{\mu}_{e})}^{4}}{(4W)^3},\\
\widetilde{E_{3}}&=& W-B{\mu}_{e},\\
\widetilde{E_{4}}& =&
-3W-\frac{{(B{\mu}_{e})}^{2}}{4W}+\frac{{(B{\mu}_{e})}^{4}}{(4W)^3}.
\eeqa

Since this example can be, in fact, solved exactly, the results
above obtained via the perturbation theory can be compared with
the exact solution of energy. In terms of the standard method,  we
immediately get the exact solution of eigenvalues and their
corresponding eigenvectors\\
\beq E_{1}^{T}=W+B\mu_{e},\, \Longrightarrow \Psi_{1}=\phi_1 \eeq
\beq E_{2}^{T}=-W+\sqrt{4W^{2}+(\mu_{e}B)^{2}},\,\Longrightarrow
\Psi_{2}=\frac{1}{\sqrt{(\omega_{42}+\omega_{42}^{T})^{2}
+4(\mu_{e}B)^{2}}}\left[(\omega_{42}+\omega_{42}^{T})\phi_2-2B\mu_{e}\phi_{4}
  \right] \eeq
\beq E_{3}^{T}=W-B\mu_{e}\; \Longrightarrow \Psi_{3}=\phi_3 \eeq
\beq E_{4}^{T}=-W-\sqrt{4W^{2}+(\mu_{e}B)^{2}},\,\Longrightarrow
\Psi_{4}=\frac{1}{\sqrt{(\omega_{42}-\omega_{42}^{T})^{2}
+4(\mu_{e}B)^{2}}}\left[(\omega_{42}-\omega_{42}^{T})\phi_2-2B\mu_{e}\phi_{4}
\right] \eeq \\
Here $\omega_{42}=E_4-E_2$, and
$\omega_{42}^{T}=E_{4}^{T}-E_{2}^{T}$.

When the magnitude of magnetic field $B$ is very weak, we can expand
the exact result of energy, which can be compared with the energy
worked out by the new way. Assuming that $B\mu_{e} \ll W$, thus \\
\beqa
E_{2}^{T} &=&-W+2W\left(1+\frac{(B\mu_{e})^{2}}{4W^{2}} \right)^{\frac{1}{2}}\nonumber \\
  &=&-W+2W\left[1+\frac{(B\mu_{e})^{2}}{8W^{2}}-\frac{(B\mu_{e})^{4}}{8(4W^{2})^{2}}+\cdots\right]\nonumber\\
                           &\approx&W+\frac{(B{\mu_{e}})^{2}}{4W}-\frac{(B\mu_{e})^{4}}{(4W)^3}\nonumber\\
                           &=&\widetilde{E_{2}}
\eeqa \beqa
E_{4}^{T} &=&-W-2W\left(1+\frac{(B\mu_{e})^{2}}{4W^{2}} \right)^{\frac{1}{2}}\nonumber \\
                       &=&-W-2W\left[1+\frac{(B\mu_{e})^{2}}{8W^{2}}
                       -\frac{(B\mu_{e})^{4}}{8(4W^{2})^{2}}+\cdots\right]\nonumber\\
                       &\approx&-3W-\frac{(B{\mu_{e}})^{2}}{4W}+\frac{(B\mu_{e})^{4}}{(4W)^3}\nonumber\\
                       &=&\widetilde{E_{4}}
\eeqa while \beq E_{1}^{T}=\widetilde{E_{1}}, \quad
E_{3}^{T}=\widetilde{E_{3}}. \eeq Obviously, we see that the
perturbed energy calculated by using Wang's scheme \cite{wang1} is
more satisfying because they have better precision than the existed
method when two kinds of results are compared with the exact
results.

\section{Calculation of IMPROVED TRANSITION PROBABILITY}\label{sec4}

In this section we will consider the transition probability from the
state $\ket{\phi_2}$ to state $\ket{\phi_4}$. First, we should
calculate its exact result. Obviously the transition probability is:
\beq P^{T}(2\to4)=\left|\bra{\phi_4}e^{-i{H}t}\ket{\phi_2}\right|^2
\eeq To work out this equation, we take use of the completeness of
the Hilbert space $\{\ket{\Psi_1}, \ket{\Psi_2}, \ket{\Psi_3},
\ket{\Psi_4}\}$ given above, which means: \beq \label{pt}
\begin{array}{lll}
P^{T}(2\to 4)
&=&\left|\sum_{\gamma_1,\gamma_2=1}^4\bra{\phi_4}\ket{\Psi_{\gamma_1}}
\bra{\Psi_{\gamma_1}}e^{-i{H}t}\ket{\Psi_{\gamma_2}}
\bra{\Psi_{\gamma_2}}\ket{\phi_2}\right|^2\\
&=&\left|\bra{\phi_4}\ket{\Psi_2}\bra{\Psi_2}
\ket{\phi_2}e^{-iE^{T}_{2}t}+\bra{\phi_4}
\ket{\Psi_4}\bra{\Psi_4}\ket{\phi_2}e^{-iE^{T}_{4}t}\right|^2\\
&=&(\mu_{e}B)^2\dfrac{\sin^2(\omega_{42}^{T}t/2)}{(\omega_{42}^{T}/2)^2}
\end{array}
\eeq where $\displaystyle
\omega^{T}_{42}=E^{T}_{4}-E^{T}_{2}=-2\sqrt{4W^{2}+(\mu_{e}B)^{2}}$.
We also write down the result given by the traditional
perturbation theory: \beq \label{p}
P(2\to4)=(\mu_{e}B)^2\dfrac{\sin^2(\omega_{42}t/2)}{(\omega_{42}/2)^2}
\eeq where $\omega_{42}=E_4-E_2=-4W$. and the result given by
Eq.(\ref{newprb}) is: \beq \label{pw}
P_{I}(2\to4)=(\mu_{e}B)^2\dfrac{\sin^2(\widetilde{\omega}_{42}t/2)}{(\omega_{42}/2)^2}
\eeq where
$\displaystyle\widetilde{\omega}_{42}=\widetilde{E}_4-\widetilde{E}_2
=-4W-\frac{{(B{\mu}_{e})}^{2}}{2W}+
2\frac{{(B{\mu}_{e})}^{4}}{(4W)^3}$. Since Eqs. (\ref{pt}),
(\ref{p}) and (\ref{pw}) have some common factors such as
$(\mu_{e}B)^2$ and $\omega_{42}^2$, in order to effectively
compare them we can let the three equations plus a common factor:
$\dfrac{(\omega_{42}/2)^2}{(\mu_{e}B)^2}$. Besides, to calculate
we should now add $\hbar$ at the right place (since we have let
$\hbar=1$ previously). That is to say, we just need to compare
$P^{T}$, $P_I$, and $P$, which are: \beq \label{it}
P^{T}=P^{T}(2\to 4)\times\dfrac{(\omega_{42}/2)^2}{(\mu_{e}B)^2}
=\dfrac{\sin^2(\omega_{42}^{T}t/2\hbar)}{(\omega_{42}^{T}/\omega_{42})^2}
=\dfrac{\sin^2{\left(\sqrt{4W^{2}+(\mu_{e}B)^{2}}t/\hbar\right)}}{1+\dfrac{(\mu_eB)^2}{4W^2}}
\eeq \beq \label{iw}
P_I=P_I(2\to4)\times\dfrac{(\omega_{42}/2)^2}{(\mu_{e}B)^2}
=\sin^2(\widetilde{\omega}_{42}t/2\hbar)=\sin^2{\left[\left(2W+\frac{{(B{\mu}_{e})}^{2}}{4W}
-\frac{{(B{\mu}_{e})}^{4}}{(4W)^3}\right)t/\hbar\right]} \eeq \beq
\label{i}P=P(2\to
4)\times\dfrac{(\omega_{42}/2)^2}{(\mu_{e}B)^2}=\sin^2(\omega_{42}t/2\hbar)
 =\sin^2{(2Wt/\hbar)}
\eeq

Next, we will replace the parameters such as $W$ and $B$ with
concrete data. From the latest data book \cite{codata}, we get the
value of $\mu_{e}=9.28476412\times10^{-24}$J$\cdot$T${}^{-1}$. Now
we explain the $W$ in Eq. (\ref{h}). Comparing with the equation
(D1) of Ref. \cite{codata}, Eq. (\ref{h}) here is written in the
form of Pauli operator but not of spin operator which is used in
(D1) of \cite{codata}. Therefore we can easily know that: \beq
W=\frac{2\pi}{\hbar}\Delta\nu_{H}\cdot\left(\frac{\hbar}{2}\right)^2=\frac{h\Delta\nu_{H}}{4}
\eeq where the $\Delta\nu_{H}$ is the so-called ground-state
hyperfine frequency, and $h$ is the Plank constant. By referring
to \cite{exp} we get the experimental value of
$\Delta\nu_{H}=1420.4057517667$ MHz. Plus the
$h=6.6260693\times10^{-34}$eV$\cdot$s (see TABEL XXV of
\cite{codata}), we finally work out the value of $W$: \beq
W=1.46858145124*10^{-6}{\rm eV} \eeq

\subsection{COMPARISON OF TRANSITION PROBABILITY AS TIME
GOES}

Now we want to compare Eqs. (\ref{it}), (\ref{i}) and (\ref{iw})
along with the change of time $t$ and magnetic field $B$
respectively. First, we compare them when time is changing, and we
set the magnitude of magnetic field as $B=10^{-3}$T, which can be
easily realized in laboratory. With all the data we get, we now
rewrite the expression of $P^{T}$, $P_I$ and $P$: \beq \label{3it}
\begin{array} {r@{\quad=\quad}l}
P^{T}&\dfrac{\sin^2(4.46320474159\times10^{9}\times
t)}{1+3.89282563044\times10^{-4}}\\[8pt]
P_I&\sin^2(4.46320474158\times10^{9}\times t)\\[8pt]
P&\sin^2(4.46233627125\times10^{9}\times t)
\end{array}\eeq

To see more clearly, we use mathematica tool to show their
differences(here we retain 12 effective figures). Note that the
function $\sin(t)$ is an oscillating one, and that the period of
(\ref{3it}) is so small that the three functions of (\ref{3it})
will oscillate intensively if the parameter $t$ varies in a large
range. This make us have to compare them within a small time span.
The graphs bellow show four small-time-spans, in which we can see
clearly their nuances(see Fig.1).
\begin{figure}[ht]
\begin{center}
\includegraphics[scale=1.0]{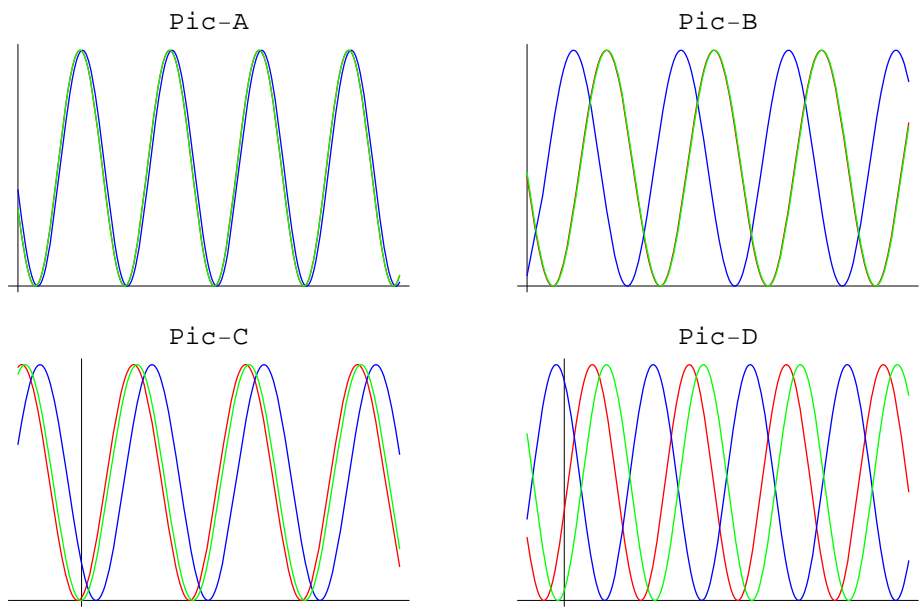}
\end{center}
\vskip -0.1in \caption{The four graphs represent four
small-time-spans, red curve represents the $P^{T}$, green the
$P_I$, and blue the $P$. We do not mark the ticks because the time
span is so small that the Mathematica tool can not even mark them
clearly,yet this does not imped us to see the differences among
the three curves of the four graphs above. The Pic-A shows while t
is around $10^{-7}$s, $P^{T}$, $P_I$ and $P$ coincide with each
other. Pic-B shows while t is around $1$s, $P^{T}$ and $P_I$ still
coincide with each other, but $P$ begins to deviate. Pic-C shows
when $t$ is around $6s$, $P_I$ begins to stray away from $P^{T}$,
while $P$ also deviates from the former two. Pic-D shows when t is
around $27.7$s, obvious deviations between $P^{T}$ and $P_I$
occur, and $P$ still shows irregular deviation.} \label{mypic1}
\end{figure}

FIG.1 tells us that when time $t$ is very small, say $10^{-7}$s, the
transition probability of traditional way and new way both coincide
with the exact solution. However, when time evolves to a relatively
large value, traditional result show irregular deviation from the
exact one, while new result still matches well with it. Of course,
new result also will deviate from the exact one when time $t$ is
large enough. Therefore we can say that, when the magnitude of
magnetic field is given, traditional result is useful only when time
$t$ is small enough, but new one will be exact in a large time-span.
As a result, we clearly see the advantage of new way of perturbation
theory. Next, we will compare the three curves when time ``$t$" is
set and magnetic field is changing.

\subsection{COMPARISON OF TRANSITION PROBABILITY AS MAGNETIC FIELD CHANGES}

At the given time we choose the time $t=1$s. We also choose four
small magnetic field spans, and FIG.2 represents the comparison of
three curves: $P^{T}$, $P_I$ and $P$. Here we write down the
expression of them:

\beq\begin{array} {r@{\quad=\quad}l}
P^{T}&\dfrac{\sin^2(\sqrt{1.991244499780254521326\times10^{19}
+7.751567612132237407735\times10^{21}\times
B^2})}{1+389.2825623868\times B^{2}}\\[8pt]
P_I&\sin^2(4.462336271259\times10^{9}+8.685548489539\times10^{11}\times{B^2}
-8.452831429358\times10^{13}\times{B^4})\\[8pt]
P&\sin^2(4.462336271259\times10^{9})
\end{array} \eeq

\begin{figure}[ht]
\begin{center}
\includegraphics[scale=1.0]{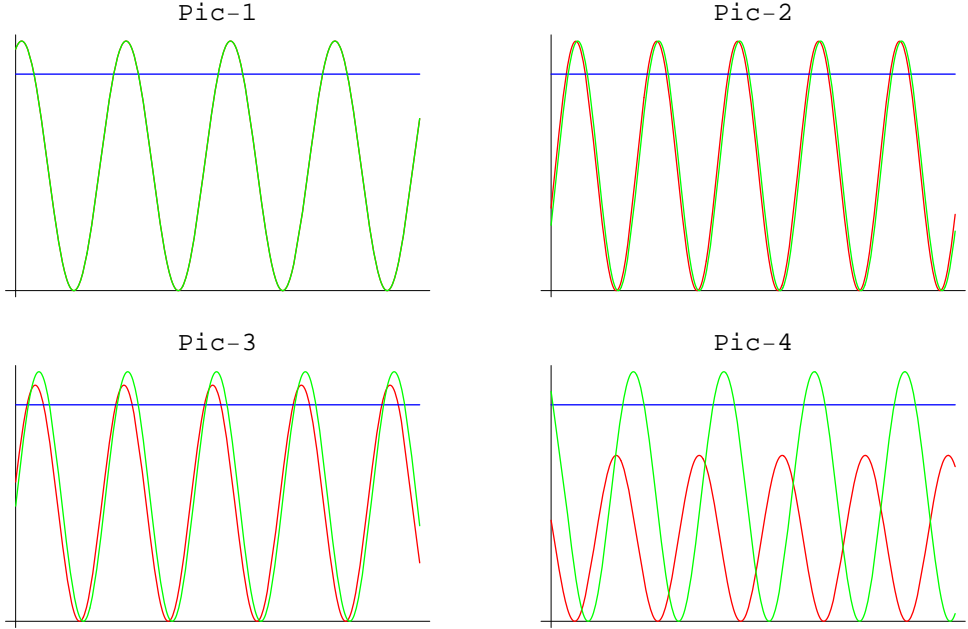}
\end{center}
\vskip -0.1in \caption{In the graph above, the red curve
represents $P^{T}$, green the $P_I$ and blue the $P$. Curve $P$ is
a straight line which is independent to $B$. Pic-1 shows when $B$
is around $10^{-4}$T, $P^{T}$ and $P_I$ coincide with each other.
Pic-2 shows when $B$ is around $1.29\times10^{-3}$T, nuance
between $P^{T}$ and $P_I$ appears. Pic-3 shows when $B$ is around
$1.21\times10^{-2}$T, there is obvious distinction between them.
Pic-4 shows when $B$ is around $0.036$T, huge difference occurs,
and even their oscillation amplitudes are different.}
\label{mypic1} \vskip -0.1in
\end{figure}

In this part, we again see the differences between traditional
result and new one. In fact, in this case the phase factor of
traditional transition probability has nothing to do with magnetic
field, which results in the straight line in FIG.2. However, as we
can see from the exact solution, the transition probability
oscillate intensively with the change of $B$, and this reveals the
defect of traditional way. Yet, new way in \cite{wang1} provide us
with a relatively satisfying expression, which also vibrates with
magnetic field ``$B$". This new expression matches well with exact
solution when ``$B$" varies in a certain range. Yet, we should
notice that there is also some flaw in the new result, which can be
seen from the Pic-4 in FIG.2. Such flaw results from the difference
of denominators between Eqs. (\ref{newprb}) and (\ref{pt}), which
will influence their vibration amplitudes.

\section{Discussion and Conclusion}\label{sec5}

Now we will come to a conclusion. In Ref. \cite{wang1}, the author
gives a improved form of perturbed solution of dynamics, it is
interesting to study its influences on some physics results. As we
can see, the phase factors correlating to dynamical behavior in the
expression of state vector are changed, and the variations absorb
the partial contributions from higher order approximations. This
leads us to consider the change of transition probability. Since
abstract discussion of the transition probability is both hard and
has no common meaning, we should find concrete cases to discuss.

Main work we do in this paper is to calculate the improved
transition probability for a Hydrogen atom in a constant magnetic
field. In fact, this example we choose is appropriate, because we
need to choose one which can be given the exact solution so that
we can compare it with the result of traditional and Wang's scheme
of perturbation theory. The comparison we give in Sec. \ref{sec4}
successfully indicate that Wang's scheme does show its advantages
beyond traditional one. Perhaps someone would argue that this
example is too simple to be meaningful enough, however, we should
note that such simple case do suggest that some expressions such
as transition probability in other cases might need amended, or at
least need perfected. Maybe such amendment does not have huge
influence in some systems, but to extend this assertion to all
problems is both too early and too cursory.

In fact, Ref. \cite{wang1} uses several skillful ways to include the
high-order approximation in the expression of low-order one, and
this procedure results in a tidy expression of perturbed solution of
dynamics. It is a physical reason why Wang's scheme can improve the
existed some conclusions. Finally, since we also see the flaw shown
in Pic-4 of Fig.2, perhaps there will still be further amendment to
it in the future.

\section*{Acknowledgments}

We are grateful all the collaborators of quantum theory group in the
institute for theoretical physics of our university. This work was
funded by the National Fundamental Research Program of China under
No. 2001CB309310, partially supported by the National Natural
Science Foundation of China under Grant No. 60573008.

\end{document}